\documentclass[10pt]{article}
\usepackage{amsfonts}
\usepackage{amssymb}
\usepackage{amsthm}
\usepackage{amsmath} 
\usepackage{graphicx}
\usepackage{hyperref}
\usepackage{enumerate}
\usepackage{color}
\usepackage[mathscr]{euscript}

\numberwithin{equation}{section}

\newcommand{\ii}{{\rm i}}
\newcommand{\ee}{{\rm e}}
\newcommand{\x}{{\rm x}}

\newcommand{\dd}{{\rm d}}
\newcommand{\beq}{\begin{equation}}
\newcommand{\ene}{\end{equation}}

\newtheorem{thm}{Theorem}

\theoremstyle{definition}

\newcommand{\ben}[1]{{\bf {\color{red} BAJ-A: }} \textcolor{red} {``#1"}} 

\begin{document}

\title{A short review of the Casimir effect with emphasis on dynamical boundary conditions}
\author{Benito A. Ju\'arez-Aubry$^1${}\thanks{Fellow Sistema Nacional de Investigadores. Email: benito.juarez@correo.nucleares.unam.mx.}\, and Ricardo Weder$^2${}\thanks{Fellow Sistema Nacional de Investigadores. Email: weder@unam.mx. Home page: http://www.iimas.unam.mx/rweder/rweder.html.}\\
 \\ \\ 
$^1$Departamento de Gravitaci\'on y Teor\'ia de Campos \\
Instituto de Ciencias Nucleares, \\ Universidad Nacional Aut\'onoma de M\'exico, \\ Apartado Postal 20-126, Ciudad de M\'exico 01000, M\'exico. \\ \\
$^2$Departamento de F\'isica Matem\'atica, \\
Instituto de Investigaciones en Matem\'aticas Aplicadas y en Sistemas, \\ Universidad Nacional Aut\'onoma de M\'exico,\\Apartado Postal 20-126, Ciudad de M\'exico 01000, M\'exico.}
\date{\today}

\maketitle

\vspace{.5cm}
\centerline{Abstract}

\noindent We give a short review on the static and dynamical Casimir effects, recalling their historical prediction, as well as their more recent experimental verification. We emphasise on the central role played by so-called {\it dynamical boundary conditions} (for which the boundary condition depends on a second time derivative of the field) in the experimental verification of the dynamical Casimir effect by Wilson et al. We then go on to review our previous work on the static Casimir effect with dynamical boundary conditions, providing an overview on how to compute the so-called local Casimir energy, the total Casimir energy and the Casimir force. We give as a future perspective the direction in which this work should be generalised to put the theoretical predictions of the dynamical Casimir effect experiments on a rigorous footing.

\section{Introduction}
\label{sec:Intro}

In the late 1940s Casimir and Polder set forth to calculate the force exerted between two polarisable atoms and between an atom and a neutral conducting plate \cite{Casimir-Polder}. Soon after Casimir understood that a force would also be present between two neutral conducting plates \cite{Casimir}. The latter effect is a particular instance of what is known today as the (static) Casimir effect. Meanwhile the former effect is related to the so-called Casimir-Polder effect.

Nowadays it is understood that the Casimir effect is explained by the non-triviality of the energy density or local energy of the ground state of a quantum system, a situation that is different to the local energy of the Minkowski vacuum in Minkowski spacetime, which can be renormalised to zero at every point. If a quantum system is confined between walls of some particular geometry, say the quantum electromagnetic field between conducting parallel plates, the non-trivial energy of the ground state will lead to a net force between the walls. This is the Casimir force. If in the previous case only one wall had been present, we would refer to the force upon it as Casimir-Polder. In general terms it is useful to keep in mind the semantics that if the system is bounded in one spatial direction we refer to the resulting net force as the Casimir effect and if it is semi-bounded we refer to it as the Casimir-Polder effect.

From a theoretical point of view, the Casimir effect is remarkable in that it is a direct way of probing properties of the ground state of a quantum system, and contributes to our conceptual understanding of what quantum field theory is all about. For this reason, it is not surprising that a large amount of effort has been devoted to studying the Casimir effect in a variety of settings, as different boundary conditions and boundary shapes can lead to attractive or repulsive forces. A couple of relatively recent reviews are \cite{Bordag, Milton}.

From a technical point of view, the relevant observable to compute is the local energy of the field, dubbed in this context the local Casimir energy, a task that requires renormalisation. Already for free theories the local energy is quadratic in the field, and thus na\"ively involves a pointwise product of distributions. There are several methods for renormalising the local Casimir energy, such as introducing a {\it high-energy regulator} (removed at the end of renormalisation), using {\it Riemann zeta regularisation} or via {\it heat kernel techniques}. The most general renormalisation method is via {\it point splitting and Hadamard subtraction}. The use of point-splitting regularisation and divergence subtraction in the context of the Casimir effect can be traced back to the works of Deutsch and Candelas, who advocated its use in curved spacetimes through general arguments \cite{Deutsch:1978sc}, and especially to the work of Kay \cite{Kay} who gave an axiomatic justification for local quantum field theories that was sufficiently robust to be applied for perturbative interacting theories and compute radiative corrections in the Casimir effect.

From an experimental viewpoint, the first verification of the Casimir effect came about around fifty years later after its prediction in an experiment carried out by Lamoreaux using an electromechanical system with two conductors and a torsion balance. The results came to an agreement of 5\% with the theoretical predictions \cite{Lamoreaux}. The first verification of the Casimir effect between metallic parallel plates took place a few years later \cite{Bressi}.

So far we have discussed the Casimir effect in the presence of fixed boundaries, but theoreticians and experimentalists have also considered a setting with moving boundaries. In this case there are analogous -- although more complicated -- effects which are collectively referred to as the {\it dynamical Casimir effect}. See e.g. \cite{Dodonov} for a review. The most impressive effect in this context is the production of particles out of the ground state by the moving boundary. The pioneering works that predicted this situation are by Moore \cite{Moore} and Davies and Fulling \cite{Davies}. The latter article is also a direct antecedent to the current vibrant field of analogue gravity (see e.g. \cite{Barcelo}), which seeks to design a measure effects in table-top experiments that emulate gravitational situations, such as superradiance or Hawking radiation. Indeed, the point of \cite{Davies} was that the creation of particles by an accelerated mirror mimicks the onset of Hawking radiation by the dynamics of spacetime during the formation of black hole (see e.g. \cite{Juarez-Aubry:2018} for details on the onset of Hawking radiation), see e.g. \cite{Juarez-Aubry:2014, Good1, Good2} also in this context.

We should mention that performing a mathematically rigorous study of the dynamical Casimir effect is challenging in general, due to the intricacies of studying (quantum) field theory in the presence of moving boundaries. We refer to \cite{Jorma, Barbado1, Barbado2} for some recent progress in studying cavities with non-uniform motion.

The experimental verification of the dynamical Casimir effect remained a challenge until recently. The reason for this is that the size of the measurable effects in amenable mechanical experimental settings is negligible, and cannot be accounted for beyond systemic errors or other sources of noise. For instance, a mechanical boundary should be accelerated to relativistic velocities in order to have a significant signal of the particle creation.

However, in 2011 the first experimental verification of the dynamical Casimir effect was reported by Wilson et al. in \cite{Wilson:2011}.\footnote{See e.g. \cite{Lahteenmaki} for other experimental measurements of the dynamical Casimir effect and \cite{Kim} for more recently proposed experiments.} In the setting of \cite{Wilson:2011} the above experimental challenges are avoided altogether. Wilson et al. considered the quantum electromagnetic field inside a waveguide that is open at one end and has at the other end attached a superconducting quantum interference device (SQUID), which induces a time-dependent magnetic flux into the waveguide changing the electromagnetic field without the need of a physical movable boundary. In other words, it is the boundary conditions imposed by the SQUID that mimic a rapidly-moving wall in the experiment. The results reported in \cite{Wilson:2011} present the detection of radiation from the open end of the waveguide in agreement with that produced by photon creation inside the waveguide due to the dynamical Casimir effect \cite{Johansson}. See \cite{Dalvit} for a non-technical review.

It turns out that a precise mathematical modelling of the above experiment is quite intricate. This owes to the fact that from a mathematical viewpoint the SQUID imposes a {\it dynamical boundary condition}, i.e., a boundary condition that depends explicitly on the second time derivative of the field at the boundary, see eq. (10) in \cite{Johansson}. This situation is unlike the better-known Dirichlet or Neumann boundary conditions that fall within the Robin class and with which a theoretical physicist is familiar.

These dynamical boundary conditions are closely related to a class of problems that in the mathematical literature are known as \emph{boundary eigenvalue problems}, see e.g. \cite{menn}. In the physics literature, systems with dynamical boundary conditions have been studied in \cite{Barbero:2015, Barbero:2017kvi, Dappiaggi:2018pju, Karabali:2015epa, Inigo, Zahn:2015due} owing to different motivations, as discussed in our papers \cite{JA-W1, JA-W2}. In the context of the Casimir effect we should also point out the work of \cite{Fosco:2013wpa}.

The purpose of the following sections will be to discuss in more detail these dynamical boundary conditions in the context of the static Casimir effect, following our work \cite{JA-W1, JA-W2}. As we shall see (eq. \eqref{Dyn} below) dynamical boundary conditions depend on a number of free parameters, which we shall take to be real constants, but we should emphasise that in the dynamical Casimir effect experiment of Wilson et al. \cite{Wilson:2011} one of these parameters is an explicit time function, which can be controlled by the experimental apparatus, and induces the electromagnetic field time variation that produces pair creation. In this sense, our work \cite{JA-W1, JA-W2} can be seen, among other things, as putting on a mathematically rigorous footing the ``in" theory describing the experimental setup of \cite{Wilson:2011} before the electromagnetic field in the waveguide begins varying in time. Another interesting aspect of systems with dynamical boundary conditions is that there is a natural way of defining ``boundary observables", which is perhaps interesting in the view of holography in high-energy physics or for condensed matter physics.

In sec. \ref{sec:Quantisation} we describe in more detail the structure of systems with dynamical boundary conditions and their quantisation, including how to define bulk and boundary observables, by studying in detail a real scalar field on an interval of (coordinate) length $\ell$ with a Robin boundary condition on the left end and a dynamical boundary condition on the right end. In sec. \ref{sec:Casimir} we study the Casimir effect for this system. We show how to obtain the local Casimir energy by a point-splitting regularisation and Hadamard renormalisation, and present the ideas on how to compute the total (integrated) Casimir energy and Casimir force with the aid of numerical techniques, which is done in more detail in \cite{JA-W2}. Perspectives are discussed in sec. \ref{sec:Perspectives}.

\section{Quantisation of systems with dynamical boundary conditions}
\label{sec:Quantisation}

Consider a real scalar field on an interval of length $\ell > 0$, $\phi: \mathbb{R} \times [0, \ell] \to \mathbb{R}$, defined by the following dynamics
\begin{align}
\label{Dyn}
\left\{
                \begin{array}{l}
                  \left[\partial^2_t- \partial^2_z + m^2 + V(z) \right] \phi(t, z) = 0, \, t \in \mathbb{R},  z \in (0,\ell),  \\
                  \cos\alpha \,\phi(t, 0) + \sin\alpha \,\partial_z \phi(t,0) = 0, \, \alpha \in [0, \pi), \\
                  \left[\beta^\prime_1 \partial^2_t -\beta_1 \right] \phi(t, \ell) = - \beta_2 \partial_z \phi(t,\ell) + \beta'_2 \partial_z \partial^2_t \phi(t,\ell),
                \end{array}\right.
\end{align}
where $m^2$ is a mass term, $V(z)$ a potential and $\beta_1, \beta_1', \beta_2, \beta_2'$ real parameters. Note that while at the end-point $z = 0$ system \eqref{Dyn} imposes a boundary condition of Robin class, the boundary condition at the endpoint $z = \ell$ depends on $\partial^2_t \phi(t, \ell)$ and $\partial_z \partial^2_t \phi(t, \ell)$, being a dynamical boundary condition in the sense discussed in the Introduction of this note. 

The form of the boundary condition indicates that $\beta^\prime_1$ can be interpreted as the square of an inverse velocity (or minus the square of an inverse velocity) of a boundary observable $\phi_\partial(t):= \phi(t,\ell)$, $-\beta_1$ yields a mass term and  $\beta_2,$ and $\beta_2^\prime$ are coupling parameters to external sources for the boundary dynamical observable. Note that the presence of normal derivatives in the sources prevents us from interpreting something like $\partial_z \phi(t, \ell)$ as a boundary observable, as these functions are not intrinsic to the boundary. The case that is relevant for the experimental verification of the dynamical Casimir effect has $\beta_2' = 0$ and is slightly more general in that $\beta_1$ is a fixed time-dependent function. Thus, system \eqref{Dyn} can be thought of as an ``in" theory in which the time dependence of $\beta_1$ has not been switched on.

If we look for solutions to \eqref{Dyn} of the form $\phi(t, z)= e^{-i\omega t}\, \varphi(z)$, \eqref{Dyn} becomes a boundary eigenvalue problem of the form
 \begin{align}
\label{w.2}
\left\{
                \begin{array}{l}
                  \left[- \partial^2_z + m^2 + V(z) \right] \varphi( z) =  \omega^2\, \varphi(z),  z \in (0,\ell),  \\
                  \cos\alpha \,\varphi(0) + \sin\alpha \,\partial_z \varphi(0) = 0, \, \alpha \in [0, \pi), \\
                  -\left[ \beta_1  \varphi( \ell)  - \beta_2 \partial_z \varphi(\ell)\right]    =  \omega^2 \left[\beta_1^\prime \varphi(\ell)  -\beta'_2 \partial_z  \varphi(\ell)\right].
                \end{array}\right.
\end{align}

Here, we have written the eigenvalue as $\lambda = \omega^2$ and note the presence of the eigenvalue in the boundary condition at $z = \ell$. The classical problem defined by eq. \eqref{w.2} has been studied by Fulton in \cite{Fulton}. In the kind of boundary eigenvalue problems like \eqref{w.2} the relevant Hilbert space where classical dynamics take place is not the standard $L^2$, but rather some extended Hilbert space including the ``boundary dynamics". Here, it is $\mathcal{H}:= L^2((0,\ell)) \oplus \mathbb{C}$ with its elements being two-component vectors, say $u = (u_1, u_2)^{\rm T}$ with $u_1 \in  L^2((0,\ell))$ and $u_2 \in \mathbb{C}$. The Hilbert space is equipped with the inner product
\begin{equation}\label{Inner}
(u,v)_\mathcal{H} := \int_0^\ell \! \dd z \, \overline{u_1(z)} {v_1(z)} +  \rho^{-1} \overline{u_2} {v_2}.
\end{equation}
where $\rho :=  \beta_1^\prime\, \beta_2 - \beta_1\, \beta_2'$ is required to be positive.

Solutions to \eqref{w.2} can be obtained by introducing the self-adjoint operator
\begin{align}
\label{opA}
\varphi \in D(A)  \mapsto A\varphi := \begin{pmatrix}
           \left[ - \partial_z^2 + m^2 + V(z) \right] \varphi_1(z)   \\
           -\left[\beta_1 \varphi_1(\ell) - \beta_2 \partial_z  \varphi_1(\ell) \right] \\
         \end{pmatrix},
\end{align}
which is densely defined in $\mathcal{H}$ on the domain
\begin{align}
\label{DomA}
D(A) & = \left\{ \varphi =  \begin{pmatrix} 
           \varphi_1   \\
           \varphi_2  \\
         \end{pmatrix} 
\in \mathcal{H}:  \varphi_1, \partial_z \varphi_1  \text{ are absolutely} \,
 \text{continuous in } [0, \ell],  \partial_z^2 \varphi_1 \in L_2((0,\ell)), \right. \nonumber \\
&\left. \cos\alpha \varphi_1(0) + \sin\alpha \partial_z \varphi_1(0) = 0, 	 \varphi_2 =  \beta_1^\prime\varphi_1(\ell) - \beta_2^\prime\partial_z \varphi_1(\ell) \right\}.
\end{align}

Indeed, the eigenvalue problem \eqref{w.2} takes the form $A \varphi = \omega^2 \varphi$ or equivalently the dynamical problem \eqref{Dyn} becomes an abstract wave equation $\partial_t^2 \phi + A \phi = 0$. Under some further technical assumptions on the coefficients $\beta_1, \beta_1', \beta_2, \beta_2'$ spelled out in \cite[Prop. 1]{JA-W1} $A$ defines a positive operator and solutions to the abstract wave equation take the usual form: Given initial data $\phi|_{t = 0} = f \in D(A)$ and $\partial_t \phi|_{t = 0} = p \in D(A^{1/2})$ strong solutions can be written in terms of a complete set of orthonormal eigenfunctions of $A$,  $\{\Psi_{n}\}_{n=1}^\infty \in D(A)$, as
\begin{align}\label{w.13}
\phi(t,z) = \sum_{n = 1}^\infty \, \Psi_{n}(z) \left[(\Psi_{n},f)_\mathcal{H} \cos(\omega_{n} t) + ( \Psi_{n},p)_\mathcal{H} \frac{\sin(\omega_{n} t)}{\omega_{n}} \right].
\end{align}

Note that the eigenfunctions $\{\Psi_{n}\}_{n=1}^\infty$ are indeed two-component vectors taking the form
\begin{align}
\Psi_n(z) = \begin{pmatrix}
           \psi_n(z)  \\
           \beta'_1 \psi_n(\ell) - \beta_2'   \partial_z\psi_n(\ell)
         \end{pmatrix},
\label{Eigenfunctions}
\end{align}
since they are in the domain of the operator $A$.

Once the classical problem has been characterised, canonical quantisation can be performed immediately. The bosonic Fock space of the theory is 
\begin{equation}
\mathscr{H} = \mathbb{C} \oplus_{n = 1}^\infty   ( \otimes_s^n \ell^2(\mathcal N)),
\label{FockSpace}
\end{equation}
where $\ell^2(\mathcal N)$, is the space of complex-valued, square-summable sequences of the form $\{ \alpha_n \}_{n=1}^\infty$, with the scalar product
\begin{equation}
\left(  \{ \alpha_n \}_{n=1}^\infty, \{ \beta_n \}_{n=1}^\infty  \right)_{l^2(\mathcal N)}:= \sum_{n=1}^\infty\, \overline{\alpha_n}\, {\beta_n},
\end{equation}
and quantum fields are operators on $\mathscr{H}$ defined in terms of the annihilation and creation operators in Fock space, $\hat a_n$ and $\hat a_n^\dagger$ resp., satisfying the canonical commutation relations. Namely, field operators are of the form  
\begin{align}\label{field}
\hat{\Phi}(t,z) = \sum_{n = 1}^\infty \frac{1}{(2 \omega_n)^{1/2}} \left(\ee^{-\ii \omega_n t} \Psi_n (z)\, \hat{a}_n + \ee^{\ii \omega_n t} \Psi_n (z)\, \hat{a}_n^\dagger \right).
\end{align}

This completes the canonical quantisation of the scalar field defined by the system \eqref{Dyn}, whereby the field observable is represented as the operator \eqref{field} in the Fock space $\mathscr{H}$ \eqref{FockSpace}. That the $3$-field operator, $\hat \varphi(z) := \hat \Phi(0,z)$, and its momentum, $\hat \pi(z) := \partial_t \hat {\Phi}(t,z)|_{t =0} $ represent canonical commutation relations follows from the commutation relations of the annihilation and creation operators.

\section{The Casimir energy and Casimir force}
\label{sec:Casimir}

The local Casimir energy can be obtained by a point-splitting prescription and Hadamard subtraction. We henceforth assume that the potential term vanishes, $V(z) = 0$. For the ground state we have that the Wightman function takes the form
\begin{align}
\langle \Omega_\ell | \hat \Phi(t,z) \, \hat \Phi(t',z') \Omega_\ell \rangle = \sum_{n = 1}^\infty \frac{1}{2\omega_n} \ee^{-\ii \omega_n(t-t')}  \Psi_n (z) \otimes {\Psi_n (z')}.
\label{G1+1}
\end{align}

Note that the two-point function is tensor-valued. On the diagonal of this tensor we have a ``bulk" and a ``boundary" two-point function, while off the diagonal we have ``bulk-boundary" two-point functions. It is natural to define a ``bulk" Casimir energy using the bulk two-point function as follows
\begin{align}
\langle \Psi |  \hat H_{\rm ren}^{\rm B}(t,z) \Psi \rangle &:= \lim_{(t',z') \to (t,z)} \frac{1}{2} \left(\partial_t \partial_{t'} + \partial_z \partial_{z'} + m^2 + V(z) \right) \left[\sum_{n = 1}^\infty \frac{1}{2\omega_n} \ee^{-\ii \omega_n(t-t')}  \psi_n (z) \otimes {\psi_n (z')} - H_{\rm M}(\x, \x') \right]. \label{CasimirEnergy}
\end{align}

Here $H_{\rm M}$ is the {\it Hadamard singular structure} of the Wightman two-point function, which for sufficiently close points\footnote{To be precise, in a normal convex neighbourhood.} takes the form
\begin{align}
 H_{\rm M}((t,z),(t',z')) & :=  -\frac{1}{4 \pi} \left\{ 2 \gamma + \ln \left[ \frac{m^2}{2} \sigma((t,z),(t',z')) \right] \right. \nonumber \\
&  + \frac{m^2}{2} \sigma((t,z),(t',z'))  \left[  \ln \left(\frac{m^2}{2} \sigma ((t,z),(t',z')) \right) + 2\gamma- 2   \right]  \nonumber \\
& \left. + \frac{m^4}{16 \pi} \sigma^2((t,z),(t',z')) \left[ \ln \left(\frac{m^2}{2} \sigma((t,z),(t',z')) \right)+2 \gamma -3 \right] \right\} \nonumber \\
&  + O\left(\sigma^3((t,z),(t',z') \ln \left( \sigma((t,z),(t',z') \right) \right), 
\label{HM}
\end{align}
where $2 \sigma((t,z),(t',z')) = -(t-t')^2+(z-z')^2$ and $\gamma$ is the Euler number. In fact, there are a number of freedoms in defining $H_{\rm M}$ but the current definition guarantees that the local energy of the Minkowski vacuum be zero. Note that in \eqref{CasimirEnergy} the ``bulk" component of the $\Psi_n$-eigenfunctions \eqref{Eigenfunctions}, $\psi_n$, is used.

The technical point in order to obtain the local Casimir energy is to estimate sufficiently well the eigenfunctions at large eigenvalue, which in the coincidence limit give rise to the distributional logarithmic divergences that are subtracted by $H_{\rm M}$, making the right-hand side of eq. \eqref{CasimirEnergy} well defined. For example, if a Dirichlet boundary condition is imposed at $z = 0$, we have that the local Casimir energy in the ground state,
\begin{align}
\Omega_\ell^{\rm (D)} = (1,0,0, \ldots),
\end{align}
is given by
\begin{align}
 \langle \Omega_\ell^{\rm (D)} |  \hat H^{{\rm B}}_{\rm ren}(t,z) \Omega_\ell^{\rm (D)} \rangle & = \frac{\pi ^2 \beta_2'+6 (2 \gamma -1) \beta_2' \ell^2 m^2+6 \beta_2' \ell^2 m^2 \ln \left(\frac{\ell^2 m^2}{4 \pi ^2}\right)+24 \beta_1' \ell}{48 \pi  \beta_2' \ell^2} \nonumber \\
 & + \frac{m^2}{2 \pi} \Re\left( \ee^{-\ii \frac{\pi}{\ell} z} \ln \left( 1- \ee^{\ii \frac{2 \pi}{\ell} z} \right) \right) + \sum_{n = 1}^\infty \left[  \left( \frac{(\mathcal{N}_n^{\rm D})^2 \omega^{\rm D}_n}{4} - \frac{\pi n }{2 \ell^2} + \frac{\pi }{4  \ell^2}  - \frac{ m^2 }{4 \pi  n} \right) \right. \nonumber \\
&  \left.  - m^2 \left( \frac{(\mathcal{N}_n^{\rm D})^2 }{8 \omega^{\rm D}_n} - \frac{1}{4 \pi n} \right) + \frac{(\mathcal{N}_n^{\rm D})^2 m^2}{4 \omega^{\rm D}_n} \sin^2\left( s_n^{\rm D} z \right) - \frac{m^2}{2 \pi n} \sin^2\left( \frac{\pi}{\ell}(n-1/2) z \right) \right],
\label{HDiriBulk}
\end{align}
where the $s_n^{\rm D}$ are connected to the eigenvalues by $(\omega_n^{\rm D})^2 = (s_n^{\rm D})^2 + m^2$ and $\mathcal{N}_n^{\rm D}$ is a normalisation factor, see \cite{JA-W1}. The ${\rm D}$ superscripts make reference to the Dirichlet boundary condition at $z = 0$. The sums on the right-hand side of  \eqref{HDiriBulk} are absolutely and uniformly convergent in $z \in (0, \ell)$. While the closed form of the eigenvalues is not known in closed form, estimations can be obtained with the aid of analytical and numerical techniques.

Note that in this case the bulk local Casimir energy is conserved in time. The local energy diverges logarithmically as $z \to 0$ and as $z \to \ell$, a feature that also occurs with two Dirichlet boundary conditions, see e.g. \cite{Fulling}. But since the logarithmic divergences of  \eqref{HDiriBulk} are integrable, the total Casimir energy, defined by
\begin{align}
E(t, \ell) := \int_{0}^\ell \! \dd z \, \langle \Omega_\ell | & \hat H^{{\rm B}}_{\rm ren}(t,z) \Omega_\ell \rangle
\end{align}
is finite. (We keep the time argument explicit out of principle, but in this case the energy is time-independent.) The Casimir force exerted on the boundaries is defined by
\begin{align}
F(t, \ell) := - \partial_\ell E(t, \ell),
\end{align}
and can also be studied with the aid of numerical techniques. We present some examples in \cite{JA-W1, JA-W2}.

In the experiments carried out by Wilson et al. \cite{Wilson:2011} the dynamical Casimir effect was studied at finite temperature at ca. $50$ mK and $250$ mK. The result \eqref{HDiriBulk} gets modified at finite temperature $T = 1/\beta$ as
\begin{align}
\langle& \hat H^{{\rm B}}_{\rm ren}(t, z) \rangle_{\beta {\rm (D)}}  = \langle \Omega_\ell^{\rm (D)} | \hat H^{{\rm B}}_{\rm ren}(t, z) \Omega_\ell^{{\rm (D)}} \rangle \nonumber \\
& + \sum_{n = 1}^\infty \left[\frac{\left( (\omega_n^{\rm (D)})^2 + m^2\right)}{2 \omega_n^{\rm (D)}} \frac{\left[\psi_n^{\rm (D)} (z)\right]^2}{\ee^{\beta \omega_n^{\rm (D)}}-1} + \frac{1}{2 \omega_n^{\rm (D)}} \frac{\left[{\partial_z\psi_n^{\rm (D)}} (z)\right]^2}{\ee^{\beta \omega_n^{\rm (D)}}-1} \right], \label{HTbulk}
\end{align}
where the sum appearing on the right-hand side of  \eqref{HTbulk} is absolutely convergent and converges exponentially fast. The above situation can also be explored in coherent states (at finite temperature), for which the general form of the local Casimir energy takes the form of the ground (thermal) state energy added to the energy of the classical solution around which the coherent state is ``peaked".

A boundary Casimir energy for the boundary observable $\phi_\partial$ can be defined too, but in this case no renormalisation is required, since in this model the boundary has spacetime dimension $1$ (i.e., the boundary has no spatial extension). 

In our paper \cite{JA-W1} the bulk and boundary Casimir energies (and state polarisations) are presented in detail for the ground state and at finite temperatures for Dirichlet and Robin (including Neumann) boundary conditions on the left end of the interval $z = 0$ and dynamical boundary conditions at $z = \ell$. This work is extended to include coherent states in \cite{JA-W2}, where the Casimir force is also explored with the aid of numerical techniques, giving strong indication that the force can be repulsive or attractive depending on the parameters of the problem. The technical part of the numerical analysis is to obtain sufficiently many ``low energy" eigenvalues, which dominate the effect by the uniform convergence of the sums, in order to obtain good numerical approximations to the Casimir force.

\section{Summary and perspectives}
\label{sec:Perspectives}

After a short discussion on the static and dynamical Casimir effects in the Introduction, we have given a short overview of the static Casimir effect for a system with dynamical boundary conditions beginning by a summary of the quantisation of systems with dynamical boundary conditions. More details appear in \cite{JA-W1, JA-W2}, including explicit numerical examples for which the Casimir force is attractive or repulsive depending on the free parameters of the dynamical boundary condition, a thorough discussion on boundary observables and details on the propagators of the theory.

We emphasise that for the experimental verification of the dynamical Casimir effect the parameter $\beta_2$ in \eqref{Dyn} is instead a time-dependent function. In this sense, the present work puts in a rigorous setting the ``in" theory that models the experiment \cite{Wilson:2011} before the electromagnetic field is set to vary in time and particle creation takes place.

The main perspective of this work is to extend the present techniques to time-dependent $\beta_2$ to recover the results in \cite{Wilson:2011, Johansson} on a rigorous footing.

\section*{Acknowledgments}
B. A. Ju\'arez-Aubry thanks the organisers of the XIX Mexican School of Particles and Fields for their kind invitation to speak at the school. B. A. Ju\'arez-Aubry is supported by a CONACYT Postdoctoral Fellowship and acknowledges additional support from project UNAM-DGAPA-PAPIIT IG100120 and CONACYT grant 140630. R. Weder acknowledges the support of project UNAM-DGAPA-PAPIIT IN100321.
\appendix

\end{document}